\documentstyle[seceq,epsf]{ptptex}

\newcommand{\beq}{\begin{equation}}
\newcommand{\eeq}{\end{equation}}
\newcommand{\beqa}{\begin{eqnarray}}
\newcommand{\eeqa}{\end{eqnarray}}
\newcommand{\kB}{\mbox{$k_{\rm B}$}}
\newcommand{\kBT}{\mbox{$k_{\rm B}T$}}

\title{
Trimer-Monomer Mixture Problem\\
 on (111) $1 \times 1$ Surface of Diamond Structure}

\author{
Noriko {\sc Akutsu}\footnote{E-mail address: nori@phys.osakac.ac.jp} 
and Yasuhiro {\sc Akutsu}$^{*,}$
\footnote{E-mail address: acts@phys.sci.osaka-u.ac.jp}
}

\inst{
Faculty of Engineering, Osaka Electro-Communication
 University\\
Neyagawa 572-8530, Japan
\\
$^*$Department of Physics, Graduate School of Science, Osaka University\\
Toyonaka 560-0011, Japan
}

\recdate{
}

\abst{
We consider a system of trimers and monomers on a triangular lattice that describes the adsorption problem on the (111) $1 \times 1$ surface of a diamond crystal structure.  We introduce a mapping to a 3-state vertex model on a square lattice.  We treat this adsorption problem using the transfer-matrix method combined with a density-matrix algorithm, to obtain thermodynamic quantities.
}

\begin{document}

\maketitle

\section{Introduction}
The two-dimensional adatom gas (adsorbate) plays important roles in a wide variety of surface phenomena.\cite{BCF,pimpinelli,desjonqueres,neddermeyer,tromp98}  For example, the growth mode or kinetic process of surface dynamics depends on the thermal adatom concentration.  Owing to recent developments of the surface observation techniques at various length scales, quantitative study of the equilibrium adatom concentration and adatom formation energy has become possible.  The thermal adatom density $c_0$ is usually given by 
\beq
c_0=n_0 \exp (- E_{af}/\kBT), \label{adatom}
\eeq
where $n_0$ is the density of atomic adsorption sites on the surface, $E_{af}$ is the adatom formation energy, $\kB$ is the Boltzmann constant, and $T$ is the temperature of the surface.  Using this formula, Tromp and Mankos\cite{tromp98} have determined the adatom formation energy for the Si(001) surface to be 0.35 $\pm$ 0.05 [eV]. (The ``adatom'' is actually a dimer, as they have also shown.)

\begin{figure}
\epsfxsize = 4 cm 
\centerline{\epsfbox{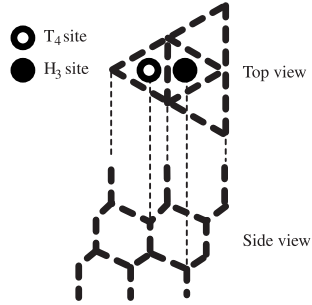}}
\caption{ Top view and side view of the (111) surface of diamond structure.}
\label{crystalst}
\end{figure}

\begin{figure}[htbp]
\epsfxsize = 10cm 
\centerline{\epsfbox{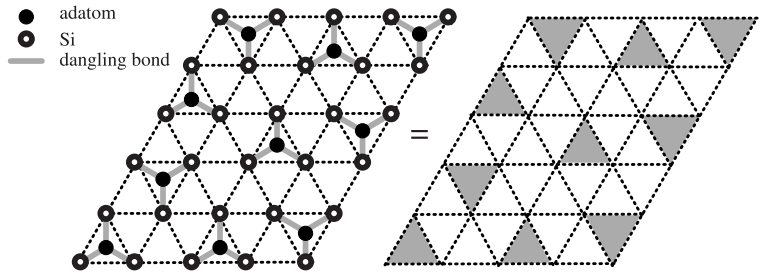}}
\caption{
Top view of the (111) surface of a diamond crystal structure with adsorbates and the corresponding triangular trimer configuration.
} \label{tile}
\end{figure}

For the Si(111) high-temperature $1 \times 1$ surface, however, the situation is somewhat more complicated, because the Si adatoms form short-range order of $\sqrt{3} \times \sqrt{3}$ symmetry.\cite{iwasaki,kohmoto,latyshev91,yang94}  From the half width of the $\sqrt{3} \times \sqrt{3}$ spot obtained by a diffraction measurement, the mean size of the ordered island is thought to be several nanometers.\cite{tromp00}
  Since the formula (\ref{adatom}) holds only in the low-density regime, where the adatoms are almost uncorrelated, it can not be applied to a system with well-developed short-range order.  Further, on the Si(111) $1 \times 1$ surface, there are two kinds of adsorption sites, called T$_4$ and H$_3$, which are common to the (111) $1 \times 1$ surface of the diamond crystal structure (Fig. \ref{crystalst}).  
 Kohmoto and Ichimiya\cite{kohmoto} showed that adatoms are adsorbed at both T$_4$ and H$_3$ sites with a mixing ratio of about 4:1, from the  analysis of the reflective high energy electron diffraction (RHEED) intensity on Si(111) $1 \times 1$ surface at 900$^{\circ}$C.  The total coverage of Si is around 0.25 monolayer (ML) (1 ML $=$ one adatom per $1 \times 1$ lattice point).

In the present paper, we consider the $1 \times 1$ phase of the (111) surface of diamond structure and make a statistical-mechanical study, taking both T$_4$ and H$_3$ sites into consideration.  In applying our analysis to the Si(111) surface, we focus on the high-temperature $1 \times 1$ phase, and our aim is to determine the relationship between the adatom concentration and the adatom formation energy.  In this sense, our present study is complementary to previous theoretical works,\cite{sakamoto,natori} in which the phase transition is the main issue. 

As shown below, our problem is basically a trimer-monomer mixture problem on a triangular lattice.  In the zero monomer density limit, it becomes equivalent to the trimer tiling (i.e. close packing) problem, whose exact solution was given recently.\cite{verberkmoes}  From the viewpoint of the tiling problem, the inclusion of monomers, i.e., removing the close-packing condition, may affect the problem in two ways: (1) the breakdown of exact solubility, and (2) the alteration of  the phase-transition-related character.   With regard to (1), we should recall that, in the well-known dimer tiling problem, the close-packing condition plays a crucial role for its Pffafian solution.\cite{dimer}  
The close-packing condition is also essential in the Bethe ansatz solution of the trimer tiling problem.\cite{verberkmoes}  With regard to (2), we should recall a rigorous proof\cite{heilmann-lieb} demonstrating the absence of the phase transition for non-zero monomer fugacity in the dimer model.  Hence, inclusion of monomers in the tiling problem is a non-trivial and interesting problem which deserves  detailed study.  In the present paper, to obtain reliable results for general cases with non-zero monomer density, we use the product wave-function renormalization group (PWFRG) method,\cite{pwfrg} which is a variant of the density matrix renormalization group (DMRG) method.\cite{dmrg}

\begin{figure}[htbp]
\epsfxsize = 9.5cm 
\centerline{\epsfbox{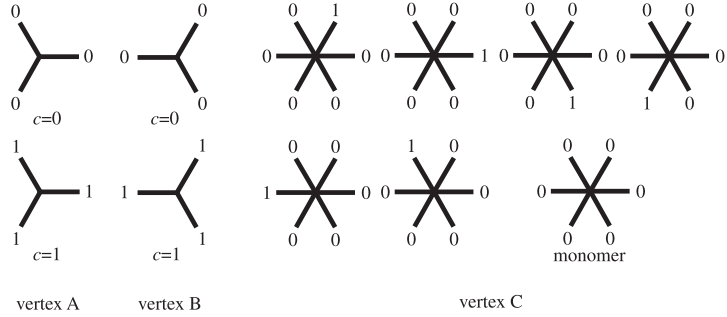}}
\caption{
Three types of elementary vertices.
}
\label{vertex1}
\end{figure}

\section{Generalized trimer tiling problem and its vertex-model representation}
We consider a triangular lattice $\Lambda$ representing the ideal (non-reconstructed) (111) surface of a diamond structure, and we place a dangling bond at each lattice point.  The adsorption sites are on its dual lattice, $\Lambda^{*}$.  We divide $\Lambda^{*}$ into two sublattices, each of which is a triangular lattice.  One sublattice represents the adsorption sites of T$_4$, and the other represents H$_3$ sites (Fig. \ref{crystalst}).  We assume that an adsorbate atom on $\Lambda^{*}$ must consume three dangling bonds around the adsorption site, and that a dangling bond is never linked to two or more adsorbates (the ``non-share'' condition) (Fig. \ref{tile}).

We can regard the adatom accompanying three dangling bonds as a ``trimer'' occupying an elementary triangle of the triangular lattice.  It should be noted that the overlap between trimers is excluded in accordance with the non-share condition.  A free dangling bond which is not linked to any adsorbate is regarded as a ``monomer'' on the triangular lattice.

\begin{figure}[htbp]
\epsfxsize = 10cm 
\centerline{\epsfbox{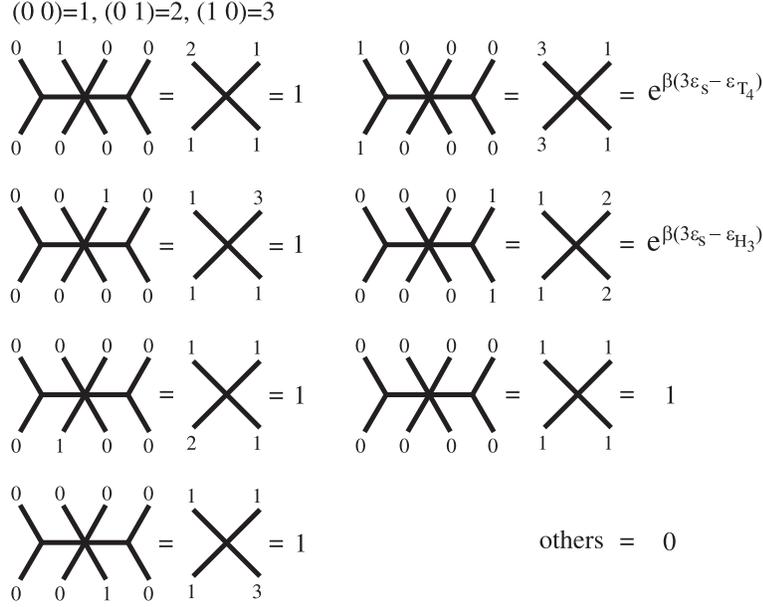}}
\caption{
Composite vertices and the three corresponding Boltzmann weights, where we make the index correspondence  $(0,0)\rightarrow 1$, $(0,1)\rightarrow 2$, and $(1,0)\rightarrow 3$.  
} \label{vertex2}
\end{figure}

The Hamiltonian of our system is   
\beq
{\cal H}= \sum_{i} (\epsilon_{{\rm T}_4} - 3\epsilon_{s}) c_{{\rm T}_4,i} + 
\sum_{i} (\epsilon_{{\rm H}_3} - 3\epsilon_{s}) c_{{\rm H}_3,i} \label{hamil},
\eeq
where the $c_{{\rm T}_4,i}$ (resp., $c_{{\rm H}_3,i}$) are the lattice gas variables at the $i$ th site of T$_4$ (resp., H$_3$).  The corresponding adsorption energy is  $\epsilon_{{\rm T}_4}$ (resp., $\epsilon_{{\rm H}_3}$). The quantity $\epsilon_{s}$ is the surface energy per $1 \times 1$ unit cell without adsorbate. The lattice-gas variable takes either the value 0 (empty) or 1 (occupied).

As is shown in Fig. \ref{vertex1}, we introduce three elementary vertices, which we denote by A, B, and C.  When a T$_4$ site is occupied by an adsorbate (resp., empty), we assign (1,1,1) (resp., (0,0,0)) to the legs of the vertex A; the same assignment applies to the vertex B for the H$_3$ site.  For the vertex C, there are seven configurations corresponding to the state of a dangling bond, as shown in Fig. \ref{vertex1}. The seventh pattern in Fig. \ref{vertex1} is regarded as the monomer in the trimer-monomer problem.

\begin{figure}[htbp]
\epsfxsize = 6cm 
\centerline{\epsfbox{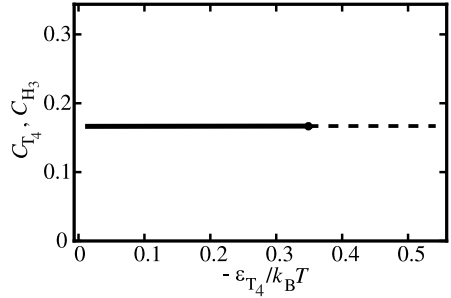}}
\caption{
Temperature dependence of $C_{{\rm T}_4}$ and $C_{{\rm H}_3}$ for a closed packed trimer problem calculated using the PWFRG method for $\epsilon_{{\rm H}_3}/\epsilon_{{\rm T}_4} = -0.5$.
The exact transition temperature is given by $ - \epsilon_{{\rm T}_4}/\kBT_m   = 4/3 \ln(3 \radical"270370{3}/4)  \approx 0.34883$.
The dashed line corresponds to the metastable state.}
\label{trimconcentration}
\end{figure}

From the elementary vertices, we build a composite 4-valent vertex.  The associated Boltzmann weights define a 3-state vertex model on the square lattice (Fig. \ref{vertex2}).  We thus have a novel mapping of the trimer-monomer mixture problem to a vertex model.

By a standard procedure,\cite{baxter} we construct the row-to-row transfer matrix ${\cal T}$ for the 3-state vertex model.  To obtain the per-site free energy from the largest eigenvalue of ${\cal T}$, we employ the PWFRG\cite{pwfrg} method to obtain an approximate (but fairly accurate) diagonalization of ${\cal T}$.  Since ${\cal T}$ is asymmetric, we apply a version of the modified PWFRG method used in Ref.~\citen{njp}, which can deal with an asymmetric transfer matrix.

\begin{figure}[htbp]
\epsfxsize = 8cm 
\centerline{\epsfbox{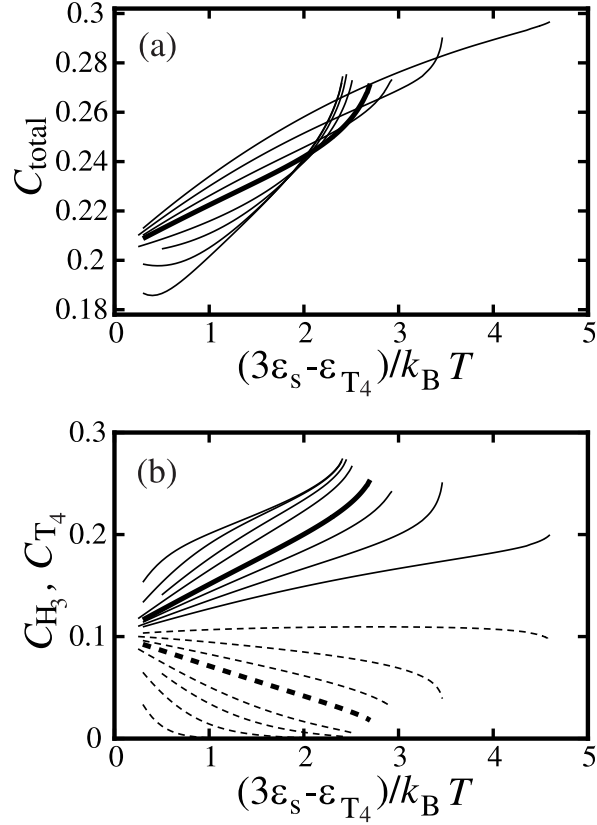}}
\caption{
Temperature dependence of concentrations calculated using the PWFRG method in the disordered phase.
(a) The quantity $C_{\rm total}$. The ratio $r=(3 \epsilon_s -\epsilon_{{\rm H}_3})/(3 \epsilon_s -\epsilon_{{\rm T}_4})$ is $0.75, 0.5, 0.25, 0, -0.5, -1, -2,-5$, from top to bottom.
 The bold curve corresponds to the case  $r=0$. 
(b) $C_{{\rm T}_4}$ (solid curves) and $C_{{\rm H}_3}$ (dashed curves). 
For $C_{{\rm H}_3}$ (resp., $C_{{\rm T}_4}$), $r=0.75, 0.5, 0.25, 0,  -0.5, -1, -2, -5$, from  top (resp., bottom) to bottom (resp., top). The bold lines correspond to the case of $r=0$.
}
\label{CvsT}
\end{figure}

\section{Calculated results}

In the limit $\epsilon_{s} \rightarrow  \infty $, the statistical weight of the monomer decreases to zero.  Then the system can be described by a close-packed trimer model whose exact solution is known.~\cite{verberkmoes}  To see the reliability of  our method, we calculate the entropy and the sublattice trimer concentrations (in the disordered phase), which we compare with the exact solution.~\cite{verberkmoes}  At $\epsilon_{{\rm T}_4}=\epsilon_{{\rm H}_3}$, the PWFRG calculation gives the entropy in the disordered phase as 0.0871$\kB$ (per $1 \times 1$ unit cell), which is in good agreement with the exact value~\cite{verberkmoes} ($(1/3) S_{\rm sym}/\kB = (1/3)\ln(3 \sqrt{3}/4) \approx 0.08720$).

\begin{figure}[htbp]
\epsfxsize = 7.0 cm 
\centerline{\epsfbox{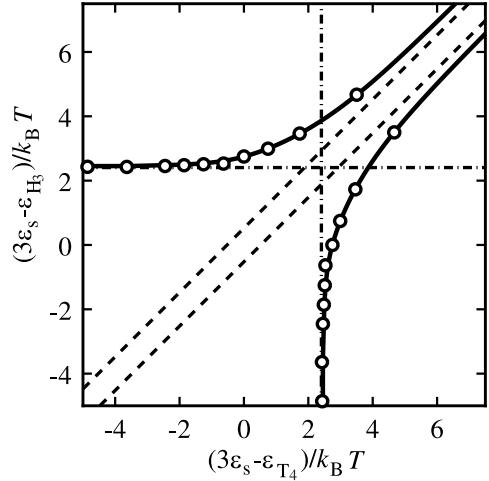}}
\caption{
Critical curves of the trimer-monomer mixtures calculated using the PWFRG method(solid curves).   Calculated critical points are denoted by the open circles. (The solid curves serve as a guide to the eye.)
The dashed lines ($(3 \epsilon_s - \epsilon_{{\rm H}_3} )/\kBT= (3 \epsilon_s - \epsilon_{{\rm T}_4} )/\kBT \pm 2 S_{\rm sym}$) correspond to the  exact transition line of the closed packed trimer problem.$^{13)}$ The dash-dotted lines correspond to  the exact critical temperature $\Delta \epsilon / \kBT_{\rm c} = \ln [(11+5 \radical"270370{5})/2] \approx 2.406$  ($\Delta \epsilon = 3 \epsilon_s -\epsilon_{{\rm H}_3}$, $3 \epsilon_s -\epsilon_{{\rm T}_4}$) of the hard hexagon model.$^{18)}$.
 }
\label{phasediagram}
\end{figure}

  In Fig. \ref{trimconcentration}, we give  the calculated result of the sublattice trimer concentrations (average number of trimers per $1 \times 1$ unit cell) $C_{{\rm T}_4}$ and $C_{{\rm H}_3}$ for $\epsilon_{{\rm H}_3}/\epsilon_{{\rm T}_4} = -0.5$. 
  Clearly, the sublattice trimer concentration $\rho$ takes a constant value $\rho_{\rm d}\sim 1/6~[\rm ML]$ in the disordered phase.  This constant behavior with $\rho_{\rm d}=1/6~[\rm ML]$ agrees with the exact solution.  
The chemical potential $\mu_{VN}$ in Ref.~\citen{verberkmoes} is given by $\mu_{VN}=(\epsilon_{{\rm T}_4}-\epsilon_{{\rm H}_3})/\kBT$.
In Ref.~\citen{verberkmoes}, it is shown that at $\mu^*_{VN} = 2 S_{\rm sym}/\kB$, the system undergoes a discontinuous order-disorder phase transition. 
Hence, in our calculation, the transition temperature $T_{\rm m}$ is given by $-\epsilon_{{\rm T}_4}/\kB T_{\rm m}=(4/3) \ln(3 \sqrt{3}/4) \approx 0.34883$.
The $\rho$-dependent free energy of Ref.~\citen{verberkmoes} has a cusp at $\rho=\rho_{\rm d}$, which suggests that the disordered phase persists as a metastable phase in a parameter region within the ordered phase near the transition line.  In our calculation, the high-temperature fixed point of the PWFRG iteration, which corresponds to the disordered phase, actually continues to exist in the ordered phase to some extent.  We thus see that our PWFRG calculation reproduces the exact solution.

For general $\epsilon_s$, the behavior of the system is governed by the ratio
\beq
r \equiv \frac{3\epsilon_s - \epsilon_{{\rm H}_3}}{3\epsilon_s - \epsilon_{{\rm T}_4}} .
\eeq
 
 The total trimer (adsorbate) concentration $C_{\rm total}$ is given by $C_{\rm total}= C_{{\rm T}_4} + C_{{\rm H}_3}$.  The monomer (free dangling bond) concentration is then given by $1-3C_{\rm total}$.  In Fig.~\ref{CvsT}, we show the temperature dependence of  $C_{{\rm T}_4}$, $C_{{\rm H}_3}$ and $C_{\rm total}$ for several values of $r$, obtained using PWFRG.  This figure gives the relationship between the thermal adsorbate concentration and the adsorption energy.  As a crosscheck, we also performed Monte Carlo calculations (not shown here).  We find that the results of these calculation agree with those of the PWFRG calculation.

In our PWFRG calculation, we assume that the system is in the disordered phase, where the maximum eigenvalue eigenstate of the transfer matrix has translational invariance.  In the ordered phase, where the translational invariance is broken and the system has period-3 structure,~\cite{verberkmoes} the PWFRG iteration ceases to converge.  Hence, if the order-disorder transition is continuous, we can determine the transition line from the  point at which the PWFRG iteration begins to lose convergence.  By such a procedure, we obtain the phase diagram of the trimer-monomer system in Fig.~\ref{phasediagram}. In the limit $r \rightarrow - \infty$ or $1/r  \rightarrow - \infty$ the system can be described by  the hard hexagon model,~\cite{baxter} whose exact transition temperature is reproduced in our calculation.

We should, however, note that to show whether the transition is continuous (second order) or discontinuous (first order) is a non-trivial problem.  In the hard hexagon limit, the transition is known to be continuous,~\cite{baxter}  while, in the trimer-tiling limit ($r\rightarrow 1$), the transition seems to be first order.  We do not discuss this point further, as it is beyond the scope of the present paper.  However, we remark here that a Monte-Carlo calculation (not shown here) suggests that, at least for the cases in which $r$ is not close to 1, the phase transition is second order, while as $r\rightarrow1$, the transition tends to exhibit a discontinuous (first-order-like) character.  Detailed discussion of such points will be given elsewhere.

\section{Application to $\mbox{ Si}$(111) $1 \times 1$ surface}

From the rocking curve of RHEED, Kohmoto and Ichimiya\cite{kohmoto} measured the adatom concentration and obtained $C_{{\rm T}_4}=0.20$, $C_{{\rm H}_3}=0.05$ and $C_{\rm total}=0.25$ at 900$^{\circ}$C.  Comparing these values with the results of our calculation, we determine the adsorption energies as $3 \epsilon_s - \epsilon_{{\rm T}_4} =0.22$ [eV/adatom] and $3 \epsilon_s - \epsilon_{{\rm H}_3}=0.056$ [eV/adatom].  The difference $\epsilon_{{\rm H}_3}- \epsilon_{{\rm T}_4}  =0.17$ eV is consistent with that estimated by Kohmoto and Ichimiya.\cite{kohmoto}

We should note here that $\epsilon_s$ depends on the ambient vapor pressure through the surface chemical potential, as the Langmuir isotherm of monolayer adsorption.\cite{desjonqueres}  Hence $\epsilon_{{\rm T}_4}- 3 \epsilon_s$ and $ \epsilon_{{\rm H}_3}- 3 \epsilon_s$ will vary with the ambient vapor pressure, which may lead to a pressure-induced phase transition. 

\section{Summary}

In the present paper, we have discussed the generalized trimer tiling problem (trimer-monomer mixture problem) on a triangular lattice, where each trimer is interpreted as a triplet of dangling bonds satisfied by a trivalent adsorbed atom.  This dangling-bond-based construction of the model, which may be regarded as an extension of the Sakamoto-Kanamori mapping\cite{sakamoto} to the hard-hexagon model, makes it directly applicable to the Si(111) $1 \times 1$ surface with both the H$_{3}$ and T$_{4}$ sites  considered as adsorption sites.  We have determined the  energy parameters of the system that consistently account for the results of experiments.  Our scheme for the construction of the tiling model is also helpful for analysis of the model; we have introduced there exists a novel mapping to a square-lattice vertex model to which we can apply the transfer-matrix method combined with the density-matrix algorithm.

By considering adsorption of the $n$-valent adatom, we can extend the present analysis to more general problems, e.g. the ``$n$-mer tiling'' problem, the ``$n$-mer and $n'$-mer mixture'' problem, and so on.  This extension is an interesting subject for future study.  Clarification of the relation to other tiling problems\cite{tiling1,tiling2,tiling3,tiling4,tiling5,tiling6,tiling7,tiling8} is also an interesting topic to be explored.  

\section*{Acknowledgements}

The authors thank Professor T. Yasue for helpful discussions.    This work was partially supported by the Research for the Future Program of the Japan Society for the Promotion of Science (JSPS-RFTF97P00201) and by a Grant-in-Aid for Scientific Research from Ministry of Education, Science, Sports and Culture (No. 12640393).

%
%
%
%
%


\end{document}